 \documentclass[preprintnumbers,amsmath,amssymb,floatfix]{revtex4}

\usepackage{graphicx}
\usepackage{dcolumn}
\usepackage{bm}
\usepackage{amssymb}
\usepackage{epsfig}
\usepackage{color}

\newcommand{\ba}{\begin{eqnarray}}
\newcommand{\ea}{\end{eqnarray}}
\newcommand{\be}{\begin{equation}}
\newcommand{\ee}{\end{equation}}
\newcommand{\bdisplay}{\begin{displaymath}}
\newcommand{\edisplay}{\end{displaymath}}

\newcommand{\eq}[1]{Eq.\,(\ref{#1})}

\begin{document}

\title{Decoupling the coupled DGLAP evolution equations: an analytic solution to pQCD}
\author{Martin~M.~Block}
\affiliation{Department of Physics and Astronomy, Northwestern University,
Evanston, IL 60208}
\author{Loyal Durand}
\affiliation{Department of Physics, University of Wisconsin, Madison, WI 53706}
\author{Phuoc Ha}
\affiliation{Department of Physics, Astronomy and Geosciences, Towson University , Towson, MD 21252}
\author{Douglas W. McKay}
\affiliation{Department of Physics and Astronomy, University of Kansas, Lawrence, KS 66045}
\date{\today}

\begin{abstract}
Using Laplace transform techniques, along with newly-developed accurate numerical inverse Laplace transform algorithms \cite{inverseLaplace1,inverseLaplace2},
we  {\em decouple} the solutions for the singlet structure function $F_s(x,Q^2)$ and $G(x,Q^2)$ of the two leading-order coupled singlet DGLAP equations, allowing us to write  fully decoupled solutions:
\ba
F_s(x,Q^2)={\cal F}_s(F_{s0}(x), G_0(x)),\nonumber\\
G(x,Q^2)={\cal G}(F_{s0}(x), G_0(x)).\nonumber
\ea
Here ${\cal F}_s$ and $\cal G$ are known functions---found using the DGLAP splitting functions---of the functions $F_{s0}(x)\equiv F_s(x,Q_0^2)$ and $G_{0}(x)\equiv G(x,Q_0^2)$, the chosen starting functions at the virtuality $Q_0^2$. As a proof of method, we  compare our numerical results from the above equations with the published MSTW LO gluon and singlet $F_s$  distributions \cite{MSTW1}, starting from their initial  values at $Q_0^2=1$ GeV$^2$. Our method completely decouples the two LO distributions, at the same time  guaranteeing that both distributions satisfy the singlet coupled DGLAP equations.  It furnishes us with a new tool for readily  obtaining the effects of the starting functions (independently) on the gluon and singlet structure functions, as functions of both $Q^2$ and $Q_0^2$. In addition, it can also be used for non-singlet distributions, thus allowing one to solve analytically for individual quark and gluon distributions values at a given $x$ and $Q^2$, with typical numerical accuracies of about 1 part in $10^5$, rather than having to evolve numerically coupled integral-differential equations on a two-dimensional grid in $x,\,Q^2$, as is currently done.
\end{abstract}


\maketitle


Accurate knowledge of gluon distribution functions at small Bjorken $x$ and large virtuality $Q^2$  plays a vital role in estimating QCD backgrounds and in calculating gluon-initiated processes, and thus in our ability to search for new physics at the Large Hadron Collider.

The gluon and quark distribution functions have traditionally been determined  simultaneously by fitting experimental data on neutral- and charged-current deep inelastic scattering processes and some jet data over a large domain  of values of $x$ and $Q^2$.  The distributions at small $x$ and large $Q^2$ are determined mainly by the  proton structure function $F_2^{\gamma p}(x,Q^2)$ measured in deep inelastic $ep$ (or $\gamma^*p$) scattering.  The fitting process starts with an initial  $Q^2_0$, typically less than the square of the  $c$ quark mass, $m_c^2\approx 2$ GeV$^2$, and  individual quark and gluon trial distributions parameterized with pre-determined shapes,  given as functions of $x$ for the chosen $Q_0^2$. The distributions are then evolved numerically on a two-dimensional grid in $x$ and $Q^2$  to larger $Q^2$ using the coupled integral-differential DGLAP equations  \cite{dglap1,dglap2,dglap3}, typically in leading order (LO) and next-to- leading order (NLO), and the results used to predict the measured quantities. The final distributions are then determined by adjusting the input parameters  to obtain a best fit to the data. This procedure is very indirect in the case of the gluon: the gluon distribution $G(x,Q^2) = xg(x,Q^2)$ does not contribute directly to the accurately determined structure function $F_2^{\gamma p}(x,Q^2)$, and is determined only through the quark distributions in conjunction with the evolution equations, or at large $x$, from jet data. For recent determinations of the gluon and quark distributions, see \cite{CTEQ6.1,CTEQ6.5,MRST,MRST4, MSTW1}.

In the following, we will summarize our analytic method for determining the singlet structure functions $F_s(x,Q^2)$ and $G(x,Q^2)$ {\em directly } and {\em individually}, using as input  $F_{s0}(x)\equiv F_s(x,Q^2_0)$ and  $G_0(x)\equiv G(x,Q^2_0)$, where $Q_0^2$ is arbitrary, with the guarantee that each individually satisfies the coupled DGLAP equations.   The method can be extended simply to embrace non-singlet functions, so that it can be used to find individual quark distributions.  However, we will not pursue that goal in this communication. Instead, we give a numerical demonstration by using the LO MSTW \cite{MSTW1} $F_{s0}$ and $G_0$ at $Q_0^2=1$ GeV$^2$  to generate their singlet structure functions and gluon distributions at large $Q^2$.   Because our basic solutions are analytic, we readily obtain numerical accuracies for $F_s(x,Q^2)$ and $G(x,Q^2)$ of better than 1 part in $10^5$ for all Bjorken-$x$ and virtuality $Q^2$ considered.

Our approach uses  a somewhat unusual application of Laplace transforms \cite{bdm1,bdm2}, in which we first introduce the variable $v\equiv \ln(1/x)$ into the coupled DGLAP equations,  then Laplace transform  these coupled integral-differential equations in $v$ space to obtain coupled {\em } homogeneous first-order differential equations in the variable $Q^2$. The parameters of these transformed  equations are known functions of $s$, the Laplace-space variable. These equations are then solved analytically. Finally, using fast and accurate numerical inverse Laplace transform algorithms \cite{inverseLaplace1,inverseLaplace2}, we transform the  solutions back into $v$ space, and, finally, into Bjorken $x$-space, i.e.,
$F_s(x,Q^2)={\cal F}_s(F_{s0}(x), G_0(x))$ and
$G(x,Q^2)={\cal G}(F_{s0}(x), G_0(x))$, where the functions $\cal F$ and $\cal G$ are determined by the splitting functions in the DGLAP equations.

Our method can be generalized to NLO, but for brevity, we will limit ourselves to LO in this paper.  We write  the coupled LO DGLAP equations  \cite{bdm1,bdm2}  as
\ba
\frac{4\pi}{\alpha_s(Q^2)}\frac{\partial F_s}{\partial\ln{Q^2}}(x,Q^2)&=& 4{F_s(x,Q^2)}+\frac{16}{3}{F_s(x,Q^2)}\ln\frac{1-x}{x}
+\frac{16}{3}x\int_x^1\left(\frac{F_s(z,Q^2)}{z}-\frac{F_s(x,Q^2)}{x}\right)\frac{dz}{ z-x}\nonumber\\
&&-\frac{8}{3}x\int_x^1F_s(z,Q^2)\left(1+\frac{x}{z}\right)\frac{\,dz}{z^2}+2n_fx\int_x^1G(z,Q^2)\left(1-2{x\over z}+2{x^2\over z^2}\right)\,{dz\over z^2},\label{dFdtauofx}\\
\frac{4\pi}{\alpha_s(Q^2)}\frac{\partial G}{\partial \ln{Q^2}}(x,Q^2)&=& {33-2n_f\over 3}{G(x,Q^2)}+12{G(x,Q^2)}\ln\frac{1-x}{x}
+12x\int_x^1\left(\frac{G(z,Q^2)}{z}-\frac{G(x,Q^2)}{x}\right)\frac{dz}{ z-x}\nonumber\\
&&+12x\int_x^1G(z,Q^2)\left({z\over x} -2 +{x\over z} -{x^2\over z^2}\right)\frac{\,dz}{z^2}+{8\over 3}\int_x^1 F_s(z,Q^2)\left(1+\left(1-{x\over z}\right)^2    \right)\,{dz\over z}.\label{dGdtauofx}
\ea
Here $\alpha_s(Q^2)$ is the running strong coupling constant, given in LO by
\be
\alpha_s(Q^2)={4\pi \over \left( 11-{2\over 3}n_f   \right)\ln(Q^2/\Lambda_{n_f}^2)},
\ee
where $\Lambda_{n_f}$ is fixed so that $\Lambda_{5}$ reproduces $\alpha_s(M_Z^2)$, and the other $\Lambda$'s ($\Lambda_4$ and $\Lambda_3$) are adjusted so that $\alpha_s$ is continuous across the boundaries $Q^2=M_b^2$ and $M_c^2$, respectively, where $M_b$ and $M_c$ are the masses of the $b$ and $c$ quarks.

We now examine the last two terms of line 1 in \eq{dFdtauofx} and rewrite them, introducing the variable changes $v=\ln (1/x)$, $w=\ln (1/z)$,  and the notation $\hat F_s(v,Q^2)\equiv F_s(e^{-v},Q^2)$,  $\hat G(v,Q^2 )\equiv G(e^{-v},Q^2)$ as
\ba
&&\frac{16}{3}{\hat F_s(v,Q^2)}\ln (e^v-1)
+\frac{16}{3}\int_0^v\left(\hat F_s(w,Q^2)-\hat F_s(v,Q^2)e^{v-w}\right)\frac{1}{e^{v-w}-1}\,dw\nonumber\\
&&=\frac{16}{3}\int_0^v{\partial \hat F_s\over\partial w}(w,Q^2)\ln \left(1-e^{-(v-w)}\right)\,dw.\label{rewrite}
\ea
where the final result---the last line in \eq{rewrite}---is found by replacing the upper limit $v$ in integral of line 1 of \eq{rewrite} by $v-\epsilon$, carrying out the integrals, doing a partial integration  and then taking the limit as $\epsilon \rightarrow 0$. Similarly, we find for the last two terms of line 1 in \eq{dGdtauofx}, that
\ba
&&12{\hat G(v,Q^2)}\ln (e^v-1)
+12\int_0^v\left(\hat G(w,Q^2)-\hat G(v,Q^2)e^{v-w}\right)\frac{1}{e^{v-w}-1}\,dw\nonumber\\
&&=12\int_0^v{\partial\hat G \over\partial w}(w,Q^2)\ln \left(1-e^{-(v-w)}\right)\,dw.\label{rewriteG}
\ea

We next rewrite \eq{dFdtauofx} and \eq{dGdtauofx}, in terms of the new variable $v$, as
\ba
\frac{4\pi}{\alpha_s(Q^2)}\frac{\partial \hat F_s}{\partial \ln{Q^2}}(v,Q^2)&=& 4{\hat F_s(v,Q^2)}
+\frac{16}{3}\int_0^v{\partial \hat F_s\over\partial w}(w,Q^2)\ln \left(1-e^{w-v}\right)\,dw\nonumber\\
&&-\frac{8}{3}\int_0^v\hat F_s(w,Q^2)\left( e^{-(v-w)}+e^{-2(v-w)}   \right)\,dw\nonumber\\
&&+2n_f\int_0^v\hat G(w,Q^2)\left(e^{-(v-w)} -2e^{-2(v-w)}+2e^{-3(v-w)}   \right)\,dw,\label{dFdtauofv}\\
\frac{4\pi}{\alpha_s(Q^2)}\frac{\partial \hat G}{\partial \ln{Q^2}}(v,Q^2)&=& {33-2n_f\over 3}{\hat G(v,Q^2)}
+12\int_0^v{\partial\hat G \over\partial w}(w,Q^2)\ln \left(1-e^{-(v-w)}\right)\,dw\nonumber\\
&&+12\int_0^v \hat G(w,Q^2)\left(1-2e^{-(v-w)}+e^{-2(v-w)}-e^{-3(v-w)} \right) \,dw\nonumber\\
&&+{8\over 3}\int_0^v\hat F_s(w,Q^2)\left( 1+\left(1-e^{-(v-w)}\right)^2            \right)\,dw.\label{dGdtauofv}
\ea
All of the integrals in \eq{dFdtauofv} and \eq{dGdtauofv} are convolutions.  Introducing Laplace transforms  allows us to factor these integrals, since the Laplace transform of a convolution is the product of the Laplace transform of the factors, i.e.,
\ba
{\cal L}\left[\int_0^v F[w]H[v-w]\,dw;s   \right]&=&{\cal L} [F[v];s]\times {\cal L} [H[v];s]\label{convolution}.
\ea
Defining the Laplace transforms
\ba
f(s,Q^2)&\equiv &{\cal L}\left[ \hat F_s(v,Q^2);s\right],\qquad
g(s,Q^2)\equiv {\cal L}[\hat G(v,Q^2);s]
\ea
and noting that
\ba
{\cal L}\left[{\partial \hat F_s \over\partial w}(w,Q^2);s\right]=s f(s,Q^2),\qquad
{\cal L}\left[{\partial \hat G \over\partial w}(w,Q^2);s\right]=s g(s,Q^2),
\ea
we can factor the Laplace transforms of \eq{dFdtauofv} and \eq{dGdtauofv} into two coupled ordinary first order differential equations in Laplace space $s$ with $Q^2$-dependent coefficients. These can be written as
\ba
{\partial f\over \partial \ln{Q^2}}(s,Q^2) &=&\frac{\alpha_s(Q^2)}{4\pi}\Phi_f (s)f(s,Q^2)+\frac{\alpha_s(Q^2)}{4\pi}\Theta_f(s)g(s,Q^2)\label{df},\\
{\partial g\over \partial \ln{Q^2}}(s,Q^2) &=&\frac{\alpha_s(Q^2)}{4\pi}\Phi_g (s)g(s,Q^2)+\frac{\alpha_s(Q^2)}{4\pi}\Theta_g(s)f(s,Q^2),\label{dg}
\ea
whose coefficients $\Phi$ and $\Theta$ are given by
\ba
\Phi_f(s)&=&4 -{8\over 3}\left({1\over s+1}+{1\over s+2}+2\left(\psi(s+1)+\gamma_E\right)\right)\label{Phif}\\
\Theta_f(s)&=&2n_f\left({1\over s+1}-{2\over s+2}+{2\over s+3} \right),\label{Thetaf}\\
\Phi_g(s)&=&{33-2n_f \over 3} +12\left({1\over s}-{ 2\over s+1}+{1\over s+2}-{1 \over s+3}-\psi(s+1)-\gamma_E\right)\label{Phig}\\
\Theta_g(s)&=&{8\over 3}\left({2\over s}-{2\over s+1}+{1\over s+2}     \right),\label{Thetag}
\ea
where $\psi(x)$ is the digamma function and $\gamma_E=0.5772156\ldots$ is Euler's constant.

The solution of the coupled equations  in \eq{df} and \eq{dg} in terms of initial values of the functions $f$ and $g$, specified as functions of $s$ at virtuality $Q_0^2$, is straightforward. The $Q^2$ dependence of the solutions is expressed entirely through the function
\be
\tau(Q^2,Q_0^2)={1\over 4 \pi}\int_{Q_0^2}^{Q^2} \alpha_s(Q'^2)\,d\,\ln Q'^2\label{tau}.
\ee
With the initial conditions $f_0(s)\equiv f(s,Q_0^2)$ and $g_0(s)\equiv g(s,Q_0^2)$, the solutions are
\ba
f(s,\tau)&= &k_{ff}(s,\tau)f_0(s)+k_{fg}(s,\tau) g_0(s),\label{f}\\
g(s,\tau)&= &k_{gg}(s,\tau)g_0(s)+k_{gf}(s,\tau) f_0(s)\label{g},
\ea
where the coefficient functions in the solution are
\ba
k_{ff}(s,\tau)&\equiv&e^{\frac{{\tau }}{2}\left(\Phi_f(s) +\Phi_g(s)\right)}\left[\cosh\left (  {\tau \over 2}R(s)\right) +\frac{\sinh\left({\tau\over2}R(s)\right)}{R(s)} \left(\Phi_f(s)-\Phi_g(s)\right)\right],\label{kff}\\
k_{fg}(s,\tau)&\equiv &\left(e^{ {\tau\over 2}\left(\Phi_f(s)+\Phi_g(s)+ R(s)\right) }-e^{ {\tau\over 2}\left(\Phi_f(s)+\Phi_g(s) - R(s)\right) }\right){\Theta_f(s)\over R(s)},\label{kfg}\\
k_{gg}(s,\tau)&\equiv &e^{{\tau \over2}\left(\Phi_f(s) +\Phi_g(s)\right)}\left[\cosh\left (  {\tau \over 2}R(s)\right) -\frac{\sinh\left({\tau\over2}R(s)\right)}{R(s)} \left(\Phi_f(s)-\Phi_g(s)\right)\right],\label{kgg}\\
k_{gf}(s,\tau)&\equiv&\left(e^{ {\tau\over 2}\left(\Phi_f(s)+\Phi_g(s)+ R(s)\right) }-e^{ {\tau\over 2}\left(\Phi_f(s)+\Phi_g(s) - R(s)\right) }\right){\Theta_g(s)\over R(s)},\label{kgf}\ea
with $R(s) \equiv \sqrt{\left(\Phi_f(s)-\Phi_g(s)\right)^2+4 \Theta_f(s)\Theta_g(s)}$.

Let us now define four  kernels $K_{FF},\ K_{FG},\ K_{GF}$ and $K_{GG}$, the inverse Laplace transforms of the $k's$, i.e.,
\ba
K_{FF}(v,\tau) &\equiv &{\cal L}^{-1}[k_{ff}(s,\tau);v],\qquad K_{FG}(v,\tau) \equiv {\cal L}^{-1}[k_{fg}(s,\tau);v],\label{fKernels}\\
K_{GG}(v,\tau) &\equiv &{\cal L}^{-1}[k_{gg}(s,\tau);v],\qquad K_{GF}(v,\tau) \equiv {\cal L}^{-1}[k_{gf}(s,\tau);v].\label{gKernels}
\ea
It is evident from Eqs.\ (\ref{tau}),  (\ref{kfg}), and (\ref{kgf}) that $K_{FG}$ and $K_{GF}$ vanish for $Q^2=Q_0^2$ where $\tau(Q^2,Q_0^2)=0$. It can also be shown without difficulty that for $\tau=0$, $K_{FF}(v,0)=K_{GG}(v,0)=\delta(v)$.

The initial boundary conditions at $Q_0^2$ are given by $F_{s0}(x)=F_s(x,Q^2_0)$ and $G_0(x)=G(x,Q^2_0)$.
In $v$-space,
$\hat F_{s0}(v)\equiv F_{s0}(e^{-v})$ and $\hat G_0(v)\equiv G_0(e^{-v})$
 are the inverse Laplace transforms of $f_{0}(s)$ and $g_0(s)$, respectively, i.e.,
\ba
\hat F_{s0}(v)&\equiv &{\cal L}^{-1}[f_0(s);v]\ {\rm and \ } \hat G_0(v)\equiv {\cal L}^{-1}[g_0(s);v].
\ea
Finally, we can write  our  {\em decoupled} solutions  in $v$-space in terms of the convolution integrals
\ba
\hat F_s(v,Q^2)&=&\int_0^v K_{FF}(v-w,\tau(Q^2,Q_0^2))\hat F_{s0}(w)\,dw +\int_0^v K_{FG}(v-w,\tau(Q^2,Q_0^2))\hat G_0(w)\,dw, \label{F}\\
\hat G(v,Q^2)&=&\int_0^v K_{GG}(v-w,\tau(Q^2,Q_0^2))\hat G_0(w)\,dw +\int_0^v K_{GF}(v-w,\tau(Q^2,Q_0^2))\hat F_{s0}(w)\,dw .\label{G}
\ea
Noting again that $v\equiv\ln(1/x)$,  in the usual variables---Bjorken-$x$ and virtuality $Q^2$---we readily find the desired {\em decoupled}   $F_s(x,Q^2)$ and $G(x,Q^2)$ from the above decoupled solutions for $\hat F_s(v,Q^2)$ and $\hat G(v,Q^2)$, requiring only a knowledge of the initial values $F_{s0}(x)$ and $G_0(x)$ at $Q_0^2$.

For {\em non-singlet} distributions $F_{ns}(x,Q^2)$, such as the difference between the $u$ and $d$ quark distributions,  $x\left[u(x,Q^2)-d(x,Q^2)\right]$, we can schematically write the logarithmic derivative of $F_{ns}$ as the convolution of $F_{ns}(x,Q^2)$ with the non-singlet splitting function ${\cal K}_{ns}(x)$ (using the convolution symbol $\otimes$), i.e.,
\ba
{4\pi\over \alpha_s(Q^2)}{\partial F_{ns}\over \partial \ln (Q^2)}(x,Q^2)&=&F_{ns}\otimes {{\cal K}}_{ns}.
\ea
After changing to the variable $v=\ln(1/x)$ and going to Laplace space $s$, we find the simple solution
\ba
f_{ns}(s,\tau)&=&e^{\tau \Phi_{ns}(s)}f_{ns0}(s),\qquad{\rm where\ } \Phi_{ns}(s)={\cal L}\left [e^{-v}\hat{\cal K}_{ns}(v);s\right ]\quad {\rm and\ }\hat{\cal K}_{ns}(v) ={\cal K}_{ns}\left(e^{-v}\right).\label{NSofs}
\ea
Thus we can find {\em any} non-singlet solution in $v$-space, using the non-singlet kernel $K_{ns}(v)\equiv {\cal L}^{-1}\left[e^{\tau \Phi_{ns}(s)};v   \right]$, by employing  the Laplace  convolution relation
\ba
F_{ns}(v,Q^2)=\int _0^v K_{ns}(v-w,\tau(Q^2,Q_0^2))\hat F_{ns0}(w)\,dw\label{Fnsofv}.
\ea
For brevity, we will not pursue the case of the non-singlet solution any further here except to note that in LO, the $\Phi_{ns}(s)$  in \eq{NSofs} is identical to $\Phi_{f}(s)$ defined in  \eq{Phif}, but will concentrate instead on the more difficult case of $F_s$ and $G$.

As an example of the application of this technique, we will compare the  $x$-space singlet distribution function $F_s(x,Q^2)$ calculated from  \eq{F}  starting from the MSTW initial conditions at $Q_0^2=1 $ GeV$^2$, with the LO MSTW \cite{MSTW1} distributions. We also will compare {\em separately} their $G(x,Q^2)$ with the results found from \eq{G}. We evaluate the kernels $K_{FF}(u,\tau),\ K_{FG}(u,\tau)$ in \eq{F} and $K_{GG}(u,\tau),\ K_{GF}(u,\tau)$ in \eq{G} numerically, using powerful new inverse Laplace transformation algorithms \cite {inverseLaplace1,inverseLaplace2}.  In order to insure continuity across the boundaries $Q^2=M_c^2$ and $M_b^2$, we will first evolve from $Q_0^2=1$ GeV$^2$ (the MSTW $Q_0^2$ value) to $M_c^2$ and use our evolved values of $G(x,M_c^2)$ and $F_s(x,M_c^2)$ for {\em new} starting values $G_0(x)$ and $F_{s0}(x)$. We will then evolve to $M_b^2$, repeating the process, thus insuring continuity of $G$ and $F_s$ at the boundaries where $n_f$ changes. We use the MSTW values  $M_c=1.40$ GeV, $M_c=4.75$ GeV, $\alpha_s(1\ {\rm GeV}^2)=0.6818$  and  $\alpha_s(M_Z^2)=0.13939$.

\begin{figure}[h]
\begin{center}
\mbox{\epsfig{file=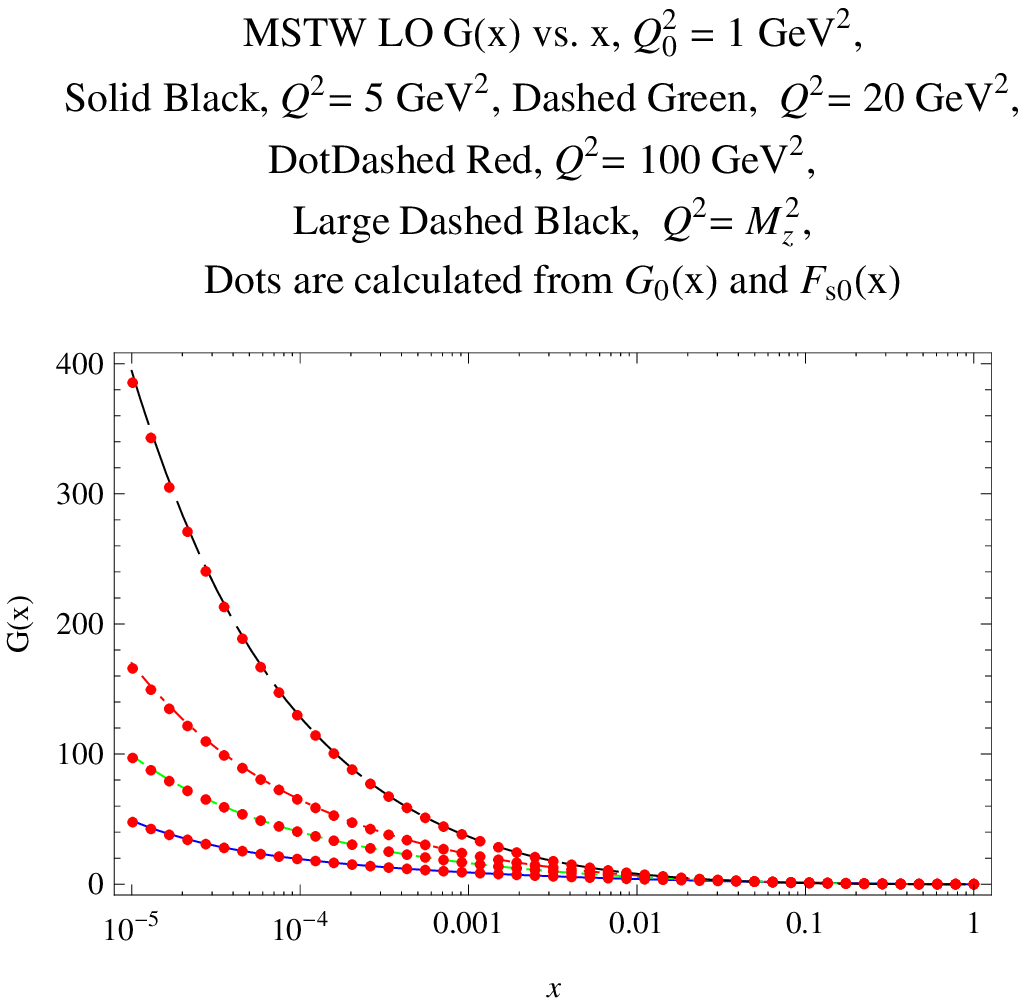
,width=5.2in%
,bbllx=0pt,bblly=0pt,bburx=317pt,bbury=190pt,clip=%
}}
\end{center}
\caption[]{
The LO MSTW \cite{MSTW1} gluon distribution, $G(x,Q^2)=xg(x,Q^2)$, for $Q^2 = 5\ ,20,\ 100$ and $M_Z^2$ GeV$^2$. The published MSTW \cite{MSTW1} curves are for $Q^2 = 5,\ 20,\ 100$ and $M_z^2$ GeV$^2$, bottom  to top. The dots are the evolution results for LO $G(x,Q^2)$ from \eq{G} (converted to $x$-space), using the LO MSTW values for $F_{s0}(x)$ and G$_0(x)$, where $Q^2_0=1$ GeV$^2$.
}
\label{fig:G}
\end{figure}

In Fig. \ref{fig:G}, we show the results---in $x$-space--- for LO $G(x,Q^2)=xg(x,Q^2)$, for 4 values of $Q^2$.  The curves are the published  MSTW \cite{MSTW1} LO gluon distributions, for $Q^2 = 5,\ 20,\ 100$ and $M_z^2$ GeV$^2$, bottom  to top. Since the MSTW collaboration \cite{MSTW1} started their evolution at  $Q_0^2=1$ GeV$^2$, we used $ F_{s0}$ and $ G_0$ constructed from their values at  $Q_0^2=1$ GeV$^2$ in  \eq{G}. The dots are our results for LO $G(x,Q^2)$ from \eq{G} (converted to $x$-space), using the LO MSTW values for $F_{s0}(x)$ and G$_0(x)$. The agreement, over this large span of $Q^2$, is quite striking. Our numerical accuracy  has been investigated and is typically better than  about 1 part in $10^5$ at small Bjorken-$x$. The most serious disagreements between our calculated $G$ and the MSTW curves are, at x= $10^{-5}$, 1.8\% and 1.6\%  at $Q^2=100$ GeV$^2$ and $M_Z^2$, respectively, which is approximately within their stated numerical errors.

\begin{figure} [h]
\begin{center}
\mbox{\epsfig{file=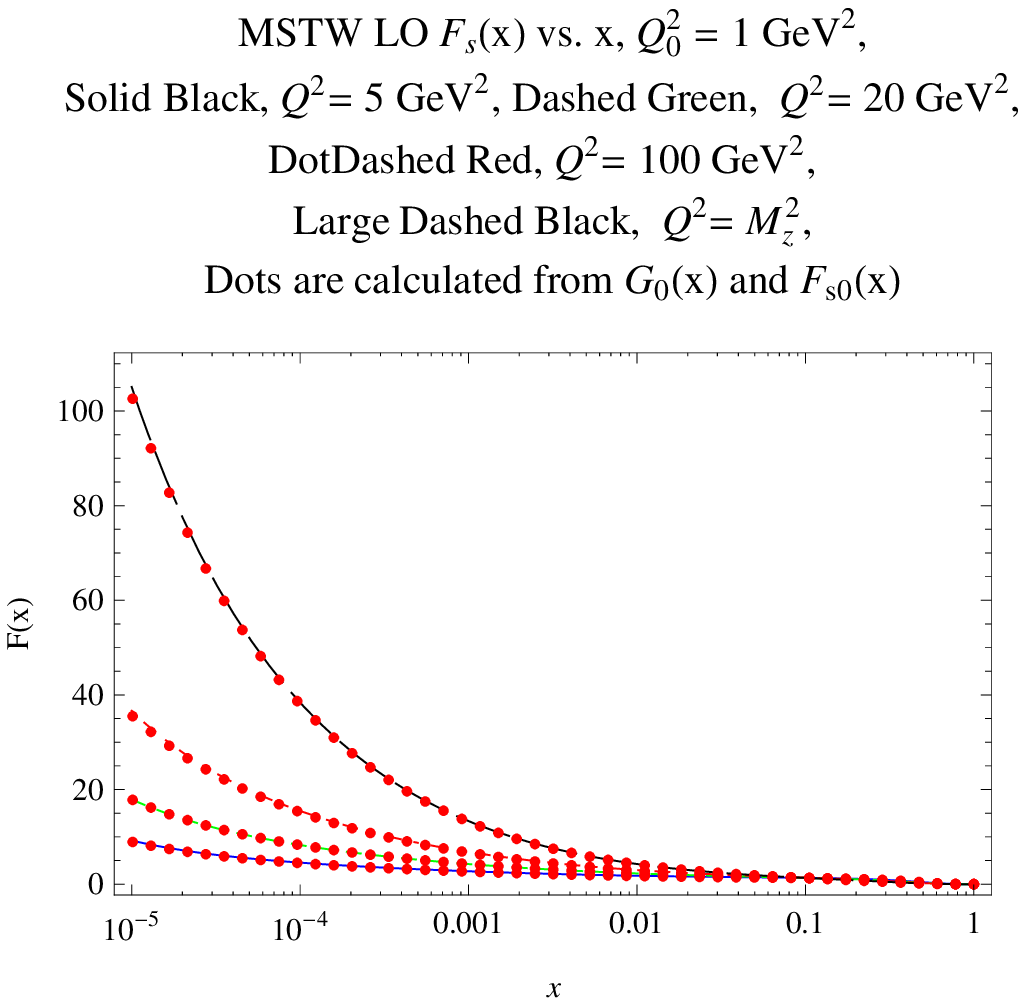
,width=5.2in%
,bbllx=0pt,bblly=0pt,bburx=317pt,bbury=190pt,clip=%
}}
\end{center}
\caption[]{
The LO MSTW \cite{MSTW1} singlet distribution, $F_s(x,Q^2)$, for $Q^2 = 5\ ,20,\ 100$ and $M_Z^2$ GeV$^2$. The published MSTW \cite{MSTW1} curves are for $Q^2 = 5,\ 20,\ 100$ and $M_z^2$ GeV$^2$, bottom  to top. The dots are the evolution results for LO $F_s(x,Q^2)$ from \eq{F} (converted to $x$-space), using the LO MSTW values for $F_{s0}(x)$ and G$_0(x)$, where $Q^2_0=1$ GeV$^2$.
}
\label{fig:F}
\end{figure}

In Fig. \ref{fig:F},  we show the results---in $x$-space--- for the
LO singlet distribution $F_s(x,Q^2)$, for 4 values of $Q^2$, where again $ F_{s0}$ and $ G_0$ are the MSTW \cite{MSTW1} values at their initial evolution value of $Q_0^2=1$ GeV$^2$. The curves are the published  MSTW \cite{MSTW1} LO singlet distributions, for $Q^2 = 5,\ 20,\ 100$ and $M_z^2$ GeV$^2$, bottom  to top. The dots are our results  for LO $F_s(x,Q^2)$ from \eq{F} (converted to $x$-space), using the LO MSTW values for $F_{s0}(x)$ and G$_0(x)$. Again, the agreement is  excellent over the entire range of Bjorken-$x$ and virtuality $Q^2$.  The most serious disagreements between our calculated $F_s$ and the MSTW curves are, at x= $10^{-5}$, 2.0\% and 1.7\%, at $Q^2=100$ GeV$^2$ and $M_Z^2$, respectively.
It is clear from Fig. \ref{fig:G} and Fig. \ref{fig:F}  for $G_s$ and $F_s$---over the enormous virtuality and $x$ span---that our analytic solutions of \eq{F} and \eq{G} are correct. The numerical values were evaluated using {\em Mathematica} \cite{Mathematica}.

In conclusion, we have constructed  {\em decoupled} analytical evolution equations for $F_s(x,Q^2)$ and $G(x,Q^2)$ from the coupled LO DGLAP equations. These require only a knowledge $F_{s0}(x)$ and $G_0(x)$, the initial values of $F_s$ and $G$ at the starting value $Q_0^2$ for the evolution, to calculate  $F_{s}(x,Q^2)$ and $G(x,Q^2)$.  The same procedures can be used for non-singlet distributions, allowing one to obtain {\em analytic} solutions for {\em individual} quark distributions, as well as  for the gluon distribution,  avoiding the necessity for numerical solutions of the coupled DGLAP on a two-dimensional grid in $x,\,Q^2$.  In essence, in a program such as Mathematica, we could define a function for each quark and gluon, and by inputting the desired $x$ and $Q^2$, simply evaluate it, accurately and rapidly.

We demonstrated numerically that the method gives  agreement with published MSTW \cite {MSTW1} LO values of $G(x,Q^2)$ and $F_s(x,Q^2)$   over an enormous range of $x$ and $Q^2$.  The accuracy obtained using our analytic solution and fast new algorithms for performing inverse Laplace transforms \cite{inverseLaplace1,inverseLaplace2} was better than 1 part in $10^5$.  In the future, as well as evaluating non-singlet distributions and the NLO singlet case,  we will evaluate $F_{s0}(x)$ and $G_{0}(x)$ in both LO and NLO, from a fit to small $x$ experimental data for the structure function $F_2^{\gamma p}(x,Q^2)$, in order to analytically obtain  accurate values of $G(x,Q^2)$ that are {\em directly } tied to experiment, which are needed for the LHC.

The authors would like to  thank the Aspen Center for Physics for its hospitality during the time parts of this work were done. P. Ha would like to thank Towson University Fisher College of Science and Mathematics for travel support.
   D.W.M. receives support from DOE Grant No. DE-FG02-04ER41308.


\bibliography{gluonsPRD.bib}

\begin{thebibliography}{13}
\expandafter\ifx\csname natexlab\endcsname\relax\def\natexlab#1{#1}\fi
\expandafter\ifx\csname bibnamefont\endcsname\relax
  \def\bibnamefont#1{#1}\fi
\expandafter\ifx\csname bibfnamefont\endcsname\relax
  \def\bibfnamefont#1{#1}\fi
\expandafter\ifx\csname citenamefont\endcsname\relax
  \def\citenamefont#1{#1}\fi
\expandafter\ifx\csname url\endcsname\relax
  \def\url#1{\texttt{#1}}\fi
\expandafter\ifx\csname urlprefix\endcsname\relax\def\urlprefix{URL }\fi
\providecommand{\bibinfo}[2]{#2}
\providecommand{\eprint}[2][]{\url{#2}}

\bibitem[{\citenamefont{Block}(2010{\natexlab{a}})}]{inverseLaplace1}
\bibinfo{author}{\bibfnamefont{M.~M.} \bibnamefont{Block}},
  \bibinfo{journal}{Eur. Phys. J. C} \textbf{\bibinfo{volume}{65}},
  \bibinfo{pages}{1} (\bibinfo{year}{2010}{\natexlab{a}}).

\bibitem[{\citenamefont{Block}(2010{\natexlab{b}})}]{inverseLaplace2}
\bibinfo{author}{\bibfnamefont{M.~M.} \bibnamefont{Block}},
  \bibinfo{journal}{private communication}
  (\bibinfo{year}{2010}{\natexlab{b}}).

\bibitem[{\citenamefont{Martin et~al.}(2009)\citenamefont{Martin, Stirling,
  Thorne, and Watt}}]{MSTW1}
\bibinfo{author}{\bibfnamefont{A.~D.} \bibnamefont{Martin}},
  \bibinfo{author}{\bibfnamefont{W.~J.} \bibnamefont{Stirling}},
  \bibinfo{author}{\bibfnamefont{R.~S.} \bibnamefont{Thorne}},
  \bibnamefont{and} \bibinfo{author}{\bibfnamefont{G.}~\bibnamefont{Watt}},
  \bibinfo{journal}{Eur. Phys. J. C} \textbf{\bibinfo{volume}{63}},
  \bibinfo{pages}{189} (\bibinfo{year}{2009}), \eprint{arXiv:0901.0002}.

\bibitem[{\citenamefont{Gribov and Lipatov}(1972)}]{dglap1}
\bibinfo{author}{\bibfnamefont{V.~N.} \bibnamefont{Gribov}} \bibnamefont{and}
  \bibinfo{author}{\bibfnamefont{L.~N.} \bibnamefont{Lipatov}},
  \bibinfo{journal}{Sov. J. Nucl. Phys.} \textbf{\bibinfo{volume}{15}},
  \bibinfo{pages}{438} (\bibinfo{year}{1972}).

\bibitem[{\citenamefont{Altarelli and Parisi}(1977)}]{dglap2}
\bibinfo{author}{\bibfnamefont{G.}~\bibnamefont{Altarelli}} \bibnamefont{and}
  \bibinfo{author}{\bibfnamefont{G.}~\bibnamefont{Parisi}},
  \bibinfo{journal}{Nucl. Phys. B} \textbf{\bibinfo{volume}{126}},
  \bibinfo{pages}{298} (\bibinfo{year}{1977}).

\bibitem[{\citenamefont{Dokshitzer}(1977)}]{dglap3}
\bibinfo{author}{\bibfnamefont{Y.~L.} \bibnamefont{Dokshitzer}},
  \bibinfo{journal}{Sov. Phys. JETP} \textbf{\bibinfo{volume}{46}},
  \bibinfo{pages}{641} (\bibinfo{year}{1977}).

\bibitem[{\citenamefont{Pumplin et~al.}(2002)}]{CTEQ6.1}
\bibinfo{author}{\bibfnamefont{J.}~\bibnamefont{Pumplin}} \bibnamefont{et~al.}
  (\bibinfo{collaboration}{CTEQ}), \bibinfo{journal}{J. High Energy Phys.}
  \textbf{\bibinfo{volume}{07}}, \bibinfo{pages}{012} (\bibinfo{year}{2002}),
  \eprint{hep-ph/0201195}.

\bibitem[{\citenamefont{Tung et~al.}(2007)\citenamefont{Tung, Lai, Belyaev,
  Pumplin, Stump, and Yuan}}]{CTEQ6.5}
\bibinfo{author}{\bibfnamefont{W.~K.} \bibnamefont{Tung}},
  \bibinfo{author}{\bibfnamefont{H.~L.} \bibnamefont{Lai}},
  \bibinfo{author}{\bibfnamefont{A.}~\bibnamefont{Belyaev}},
  \bibinfo{author}{\bibfnamefont{J.}~\bibnamefont{Pumplin}},
  \bibinfo{author}{\bibfnamefont{D.}~\bibnamefont{Stump}}, \bibnamefont{and}
  \bibinfo{author}{\bibfnamefont{C.-P.} \bibnamefont{Yuan}},
  \bibinfo{journal}{J. High Energy Phys.} \textbf{\bibinfo{volume}{02}},
  \bibinfo{pages}{053} (\bibinfo{year}{2007}), \eprint{hep-ph/0611254}.

\bibitem[{\citenamefont{Martin et~al.}(2002)\citenamefont{Martin, Roberts,
  Stirling, and Thorne}}]{MRST}
\bibinfo{author}{\bibfnamefont{A.~D.} \bibnamefont{Martin}},
  \bibinfo{author}{\bibfnamefont{R.~G.} \bibnamefont{Roberts}},
  \bibinfo{author}{\bibfnamefont{W.~J.} \bibnamefont{Stirling}},
  \bibnamefont{and} \bibinfo{author}{\bibfnamefont{R.~S.}
  \bibnamefont{Thorne}}, \bibinfo{journal}{Eur. Phys. J. C}
  \textbf{\bibinfo{volume}{23}}, \bibinfo{pages}{73} (\bibinfo{year}{2002}),
  \eprint{hep-ph/0110215}.

\bibitem[{\citenamefont{Martin et~al.}(2004)\citenamefont{Martin, Roberts,
  Stirling, and Thorne}}]{MRST4}
\bibinfo{author}{\bibfnamefont{A.~D.} \bibnamefont{Martin}},
  \bibinfo{author}{\bibfnamefont{R.~G.} \bibnamefont{Roberts}},
  \bibinfo{author}{\bibfnamefont{W.~J.} \bibnamefont{Stirling}},
  \bibnamefont{and} \bibinfo{author}{\bibfnamefont{R.~S.}
  \bibnamefont{Thorne}}, \bibinfo{journal}{Phys. Lett. B}
  \textbf{\bibinfo{volume}{604}}, \bibinfo{pages}{61} (\bibinfo{year}{2004}),
  \eprint{hep-ph/0410230}.

\bibitem[{\citenamefont{Block et~al.}(2008)\citenamefont{Block, Durand, and
  McKay}}]{bdm1}
\bibinfo{author}{\bibfnamefont{M.~M.} \bibnamefont{Block}},
  \bibinfo{author}{\bibfnamefont{L.}~\bibnamefont{Durand}}, \bibnamefont{and}
  \bibinfo{author}{\bibfnamefont{D.~W.} \bibnamefont{McKay}},
  \bibinfo{journal}{Phys. Rev. D} \textbf{\bibinfo{volume}{77}},
  \bibinfo{pages}{094003} (\bibinfo{year}{2008}), \eprint{arXiv:0710.3212
  [hep-ph]}.

\bibitem[{\citenamefont{Block et~al.}(2009)\citenamefont{Block, Durand, and
  McKay}}]{bdm2}
\bibinfo{author}{\bibfnamefont{M.~M.} \bibnamefont{Block}},
  \bibinfo{author}{\bibfnamefont{L.}~\bibnamefont{Durand}}, \bibnamefont{and}
  \bibinfo{author}{\bibfnamefont{D.~W.} \bibnamefont{McKay}},
  \bibinfo{journal}{Phys. Rev. D} \textbf{\bibinfo{volume}{79}},
  \bibinfo{pages}{014031} (\bibinfo{year}{2009}), \eprint{arXiv:0808.0201
  [hep-ph]}.

\bibitem[{Mat(2009)}]{Mathematica}
\bibinfo{journal}{{\em Mathematica} 7, a computing program from Wolfram
  Research, Inc., Champaign, IL, USA, www.wolfram.com}  (\bibinfo{year}{2009}).

\end{thebibliography}

\nopagebreak

\end{document}